\documentclass[11pt]{article}
\usepackage[a4paper, margin=1in]{geometry}
\usepackage{authblk}         
\usepackage{amsmath, amssymb}
\usepackage{graphicx}
\usepackage{authblk}
\usepackage{hyperref}
\usepackage{natbib}          
\usepackage{doi}             

\title{ShortCake: An integrated platform for efficient and reproducible single-cell analysis}

\author[1]{Ryuichiro Nakato\thanks{Corresponding author: \texttt{rnakato@iqb.u-tokyo.ac.jp}}}
\author[1]{Luis Augusto Eijy Nagai}
\affil[1]{Laboratory of Computational Genomics, Institute for Quantitative Biosciences, The University of Tokyo, Tokyo 113-0032, Japan.}

\date{\today}

\begin{document}

\maketitle

\begin{abstract}
Motivation: Recent advances in single-cell analysis have introduced new computational challenges. Researchers often need to use multiple analysis tools written in different programming languages while managing version conflicts between related packages within a single workflow. For the research community, minimizing the time spent on environment setup and installation issues is essential. \\\
Results: We present ShortCake, a containerized platform that integrates a suite of single-cell analysis tools written in R and Python. ShortCake isolates competing Python tools into separate virtual environments that can be easily accessed within a Jupyter notebook. This enables users to effortlessly transition between various environments, including R, even within a single notebook.
Additionally, ShortCake offers multiple ``flavors,'' enabling users to select container images tailored to their specific needs. ShortCake provides a unified environment with fixed versions of various tools, thus streamlining workflows, reducing setup time, and improving reproducibility.\\\
Availability and implementation: The ShortCake image is available on DockerHub\\\ (https://hub.docker.com/r/rnakato/shortcake). The source code is available on GitHub (https://github.com/rnakato/ShortCake).
\end{abstract}

\section{Introduction}

Single-cell analysis is a powerful method for exploring cellular heterogeneity, lineage, and spatial information by profiling the transcriptome and epigenome of individual cells (\cite{Lareau2019, Stuart2019}). 
The landscape of single-cell analysis is rapidly expanding. Currently, even for single-cell RNA sequencing (scRNA-seq) alone, more than 1,800 tools are registered in the scRNA-tools database (\cite{Zappia2018}). 
As a result, it has become common to combine multiple tools within a single project, or to test and compare different methods for the same analytical step, while working in multiple programming languages. 
Managing these heterogeneous environments can pose significant challenges, especially for researchers without a bioinformatics background.

Several pipelines have been developed to facilitate such integrated single-cell data analysis using multiple tools.
For instance, the Seurat ecosystem in R and the scverse in Python offer valuable frameworks for single-cell analysis (\cite{Seuratv5, Virshup2023}). However, these ecosystems do not provide environments for both R and Python, and do not solve installation problems.
Users often struggle with dependency conflicts, version mismatches, and complex setup procedures when trying to combine various tools for their own custom analysis workflows. Although package managers like Conda (https://anaconda.org/) are helpful, they still have difficulty reconciling conflicting package dependencies. These problems can prevent researchers from carrying out their research ideas.

Another challenge is reproducibility, especially across different computational environments. For instance, identical workflows have produced different sets of detected transcripts on macOS versus Linux (\cite{DiTommaso2017}). Improving the reproducibility of previous studies has been a key challenge.

To address these challenges, we developed ShortCake, a Docker-based platform that consolidates various single-cell RNA-seq and ATAC-seq (scATAC-seq) analysis tools in both R and Python environments, as well as correlated command-line tools. 
ShortCake separates conflicting Python packages into separate virtual environments, each of which is accessible via a Jupyter Notebook kernel. This streamlines the process of switching between tools without exiting an interactive session. 
ShortCake also provides multiple ``flavors'' of the image, allowing users to download only the necessary components and conserve computational resources. Additionally, since the ShortCake Dockerfile is publicly available, users can also edit it to build custom images that include any necessary additional tools.
This architecture dramatically reduces installation costs and facilitates reproducibility among users and host computers. It lowers the barrier to entry for researchers and promotes reproducibility through standardized environments. 

\section{Design and Implementation}

\subsection{Overall architecture}
Shortcake is distributed as a Docker image built on an Ubuntu 22.04 base layer. Docker packages software and its dependencies into lightweight containers, ensuring that the same environment runs identically on any host computer. To enable GPU computation, the latest version of Shortcake (v3.3.0) uses the CUDA 11.8.0 runtime and cuDNN 8 for Ubuntu 22.04.
R is installed directly in the container image (v3.3.0 ships with R 4.4.1). Python environments are managed with Micromamba (https://github.com/mamba-org/micromamba-releases), and each package in the base environment is version-pinned via an env.yaml file.

ShortCake (v3.3.0) includes over 90 tools for single-cell analysis. These tools cover various steps, including quality control, doublet detection, batch integration, trajectory/velocity inference, spatial analysis, multimodal integration, and network reconstruction. ShortCake also includes reference genomes, gene annotations, and related demo datasets. See Supplementary Table 1 for a complete list.

\subsection{Jupyter Notebook execusion}
The recommended workflow is to start ShortCake's Docker container, launch Jupyter Notebook, and connect through a web browser, for example:

\begin{verbatim}
docker run --rm -p 8888:8888 rnakato/shortcake jupyternotebook.sh
\end{verbatim}

Users can also run the container on a remote server and access it from their local laptop.

ShortCake resolves conflicts among Python-based tools by assigning each tool to its own virtual environment. Tools that invoke other tools internally are bundled into the same environment. For instance, UnitVelo (\cite{Gao2022}), which depends on scvi-tools (\cite{Gayoso2022}), shares the same virtual environment.

\begin{figure}
    \centering
    \includegraphics[width=1\linewidth]{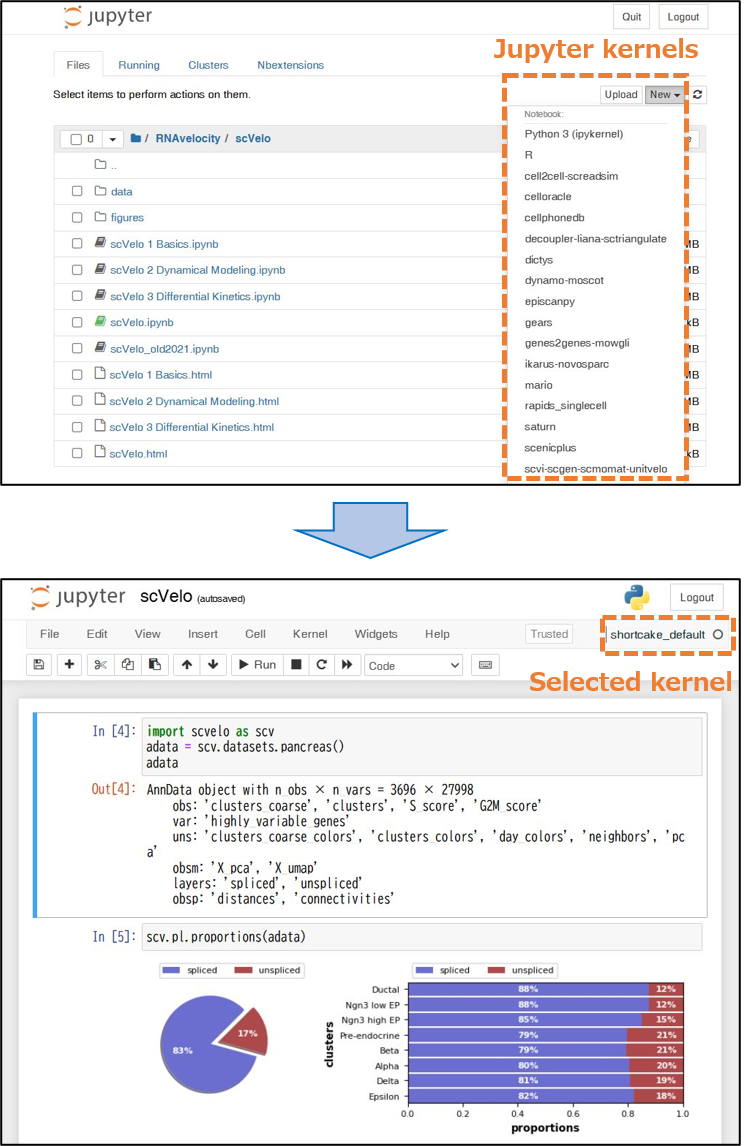}
    \caption{The Jupyter notebook launched from the Shortcake image. Top: Users can select the desired kernel for each virtual environment when creating a new notebook. Bottom: The selected kernel (virtual environment) is activated in the notebook.}
    \label{fig:jupyter_kernel}
\end{figure}

To streamline access to the many environments, ShortCake registers a dedicated Jupyter kernel for each, making them selectable within the Jupyter Notebook (Figure~\ref{fig:jupyter_kernel}). This design enables users to effortlessly switch between environments within a single notebook and facilitates workflows that rely on multiple tools with conflicting dependencies. By clicking the ``New'' button in Jupyter Notebook and selecting the desired kernel, users can launch an analysis session that runs inside their chosen virtual environment.

\subsection{Rstudio execusion}
Although the R environment can also be invoked in Jupyter Notebook, certain R-based tools do not run properly in the interface (e.g., the interactive functions of Monocle3 (\cite{Qiu2017})). 
Shortcake can launch RStudio (https://docs.posit.co/ide/user/) with this command:

\begin{verbatim}
docker run --rm -p 8888:8888 rnakato/shortcake rstudio
\end{verbatim}
so that users can begin their analyses in the RStudio environment, as all of the R libraries in Shortcake are already installed.

\subsection{Command-line excusion}
Several single-cell tools provide command-line tools. For example, Velocyto (\cite{Velocyto}) provides the command \texttt{velocyto run10x} to generate a .loom file.
It can be executed as follows:

\begin{verbatim}
docker run --rm -p 8888:8888 rnakato/shortcake \
velocyto run10x -m repeat_msk.gtf mypath/genes.gtf
\end{verbatim}

It is also possible to log directly into the ShortCake container and work inside it using the command-line interface.

\begin{verbatim}
docker run --rm -p 8888:8888 rnakato/shortcake /bin/bash
\end{verbatim}

\subsection{ShortCake flavors}\label{sec:flavors}
One drawback of ShortCake is that the Docker image grows rapidly when it contains many virtual environments. Deep-learning tools that require CUDA libraries tend to be particularly large, often adding more than 10 GB to a single virtual environment. Consequently, the ShortCake image with the full model exceeds 100 GB on Docker, which is impractical for most laptops.

To circumvent this problem, we have prepared several ``flavors'' of ShortCake. By providing lightweight images that only include the set of tools most users need, it is easier to use for purposes such as tutorials.
The main flavors are outlined below:

\begin{itemize}
  \item \texttt{shortcake\_seurat}: Contains only Seurat (\cite{Seuratv5}) and its related packages.
  \item \texttt{shortcake\_r}: Builds on \texttt{shortcake\_seurat} with additional R packages. Jupyter Notebook is available, but no Python single-cell tools are installed.
  \item \texttt{shortcake\_light}: Adds the base Python environment to \texttt{shortcake\_r}. This flavor bundles Seurat, Scanpy (\cite{Scanpy}), Monocle3 (\cite{Qiu2017}), and scVelo (\cite{scVelo}). This configuration is sufficient for most users.
  \item \texttt{shortcake}: Extends \texttt{shortcake\_light} with nearly all remaining Python virtual environments.
  \item \texttt{shortcake\_full}: The comprehensive image, including every supported tool.
\end{itemize}

\subsection{Customization and extensibility}
Users can extend ShortCake either by modifying its original Dockerfile or by creating a new Dockerfile that begins with 'FROM rnakato/shortcake'.
This flexibility enables users to incorporate the latest tools and their own scripts, making it easy to create custom analysis pipelines with ShortCake.

\section{Discussion}

ShortCake has been continuously updated since its inception in 2022 and has been utilized in several studies (\cite{Nagai2023, Shibata2024}).
Its Docker-based approach enables users to replicate an identical analysis environment on any local machine with minimal effort, ensuring reproducible workflows. When Docker privileges are unavailable (e.g., on shared cluster servers), the same image can be run with Singularity (\cite{Singularity}) instead.

Community initiatives, such as nf-core (\cite{nf_core_scrnaseq}), have successfully standardized single-cell workflows. These initiatives provide curated Nextflow pipelines that can be executed reproducibly. This type of framework is ideal for large consortia that must run a single, agreed-upon pipeline on numerous samples in a tightly controlled environment.
Conversely, ShortCake is designed for individual researchers or small groups exploring suitable case-specific workflows by iterating over alternative tools, parameter settings, and custom scripts. Rather than enforcing a fixed workflow, ShortCake offers a flexible interface that accelerates the prototyping phase for specific biological questions.

As the single-cell research field continues to advance, new technologies will emerge, such as spatial and perturbation analyses. We aim to keep up with these advances by providing the research community with well-validated tools that will facilitate their work.

\section{Competing interests}
No competing interest is declared.

\section{Author contributions}

R.N. developed ShortCake. R.N. and L.A.E.N. maintained and tested it. R.N. and L.A.E.N. wrote and revised the manuscript. 

\section{Acknowledgments}
This work was supported by a Grant-in-Aid for Scientific Research under grant number 23H02466, the Japan Agency for Medical Research and Development under grant number JP23gm6310012h0004, and the JST FOREST Program under grant number JPMJFR224Y.

\bibliographystyle{plainnat}
\bibliography{references}

\end{document}